# Multi-vortex dynamics in junctions of charge density waves

T. Yi[a], A. Rojo Bravo[b], N. Kirova [c,d,1] and S. Brazovskii[b,d]

[a] *South University of Science and Technology of China, Shenzhen, Guangdong 518055, China*
[b] *CNRS, LPTMS, URM 8502, Univeristé Paris-sud, Orsay, 91405, France*
[c] *CNRS, LPS, URM 8626, Univeristé Paris-sud, Orsay, 91405, France*
[d] *International Institute of Physics, 59078-400 Natal, Rio Grande do Norte, Brazil*

**Abstract**
Ground state reconstruction by creation of topological defects in junctions of CDWs is a convenient playground for modern efforts of field-effect transformations in strongly correlated materials with spontaneous symmetry breakings. Being transient, this effect contributes also to another new science of pump-induced phase transitions. We present a dynamical model for behavior of the CDW in restricted geometries of junctions under an applied voltage or a passing current. The model takes into account multiple interacting fields: the amplitude and the phase of the CDW complex order parameter, distributions of the electric field, the density and the current of various normal carriers. A particular challenge was to monitor the local conservation of the condensed and the normal charge densities. That was done easily invoking the chiral invariance and the associated anomaly, but prize is an unconventional Ginsburg-Landau type theory which is not analytic with respect to the order parameter. The numerical modeling poses unusual difficulties but still can demonstrate that vortices are nucleated at the junction boundary when the voltage across, or the current through, exceed a threshold.

## 1. Introduction

Inhomogeneous and transient states of cooperative electronic systems with a spontaneous symmetry breaking attract a persistent attention: from vortices in superconductors to stripes in doped oxides, charge ordering in organic conductors, to recent local probes in charge density waves (CDW), and to latest studies of transformations under impacts of a high electric field or pulses of light. The most natural background for these effects is provided by electronic crystals which notion generalizes all cases of special organization of charges in conducting solids, see [1-3]. A unique property of electronic crystals is related to the possibility of the collective current conduction by sliding which is ultimately related to appearance of inhomogeneous superstructures under stresses from the electric field.

Being a crystal of singlet electronic pairs, the CDW is a particular form of an electronic crystal which is most accessible experimentally and best treatable theoretically. In the CDW ground state, the elementary units can be readjusted by absorbing or rejecting pairs of electron. Such a phase-slip process should go via topologically nontrivial configurations: solitons and dislocations – the CDW vortices [4,5]. A new experimental access to those effects came from studies of nano-fabricated mesa-junctions [6,7], from the STM visualizations [8], and from the X-ray micro-diffraction [9,10]. A static structure of topological defects can be formed in the CDW under applied transverse voltage [11,12] or current or under a surface strain [13] or injection [14].

Recently [15-17] we performed a program of modeling the stationary states and their transient dynamic for the CDW in restricted geometries of junctions under the applied field or the passing current. The model exploited dissipative Ginzburg-Landau (GL) type equation for the complex order parameter $\Psi = A\exp(i\varphi)$ augmented to take into account other interacting quantities: the electric field, the density $n_{ex}$ and the $n_{ex}$ current $j_{ex}$ of normal carriers. The explicitly treated normal carriers were only the extrinsic ('*ex*') ones, which do not interact with CDW directly, but only via the common electric potential $\Phi$. The intrinsic ('*in*') carriers (occupying proximity of the CDW gap in the momentum space) were supposed, in a spirit of the GL approach, to be integrated out and enter to the model only via temperature dependent parameters. This model looked well justified and the numerical procedures worked very well. Nevertheless, this approach has some disadvantages: i. It is not convenient for considering the kinetics in a gapful charge density wave where the intrinsic carriers dominate while the extrinsic ones may not be even present; ii. It makes

---

difficult to address the current conversion (between the normal carriers and the condensed electrons) when the particles balance must be carefully monitored; iii. Most drastically, it does not allow to monitor the local charge conservation in case of strong time *t* and space *r* dependence of both the amplitude $A(r,t)$ and the phase $\varphi(r,t)$ which just takes place at cores of moving vortices as discussed below.

In this article we shall derive the extended GL-like theory for the CDW which allows taking intrinsic carriers explicitly into account and preserve the charge conservation locally. Finally, we shall present some results from numerical studies of this model showing a nucleation of electronic vortices (CDW dislocations).

## 2. Extended GL model for CDW from chiral invariance and anomaly.

In a CDW system there may be two types of normal carriers with concentrations $n_{in}$ and $n_{ex}$ for intrinsic and extrinsic ones, see Fig.1. Extrinsic carriers belong to other bands and/or to open parts of the Fermi surface which are distant from the CDW forming areas; they are coupled with the CDW only via the Coulomb potential. Intrinsic carriers interact also with the CDW deformation – compressing strain $q=\varphi'=\partial_x\varphi$. For the intrinsic carriers, the electronic spectrum is formed by the CDW gap and their energies move up and down when the Fermi level breathes as $\delta E_F=(\hbar v_F/2)q$ which affects their density. While they need to be activated across the gap, $n_{in}$ are present in all CDW materials. Among known quasi-1D CDW compounds, extrinsic carriers appear in NbSe$_3$. Many quasi-2D CDW compounds (see [3]) keep non-gaped parts of the Fermi surface because of their incomplete nesting.

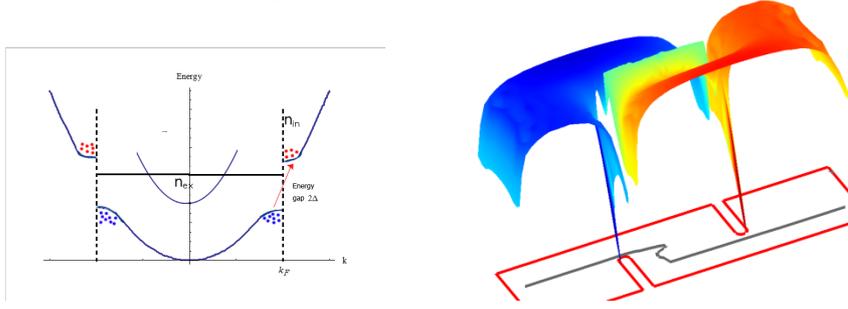

Fig1. Left panel: Electrons' spectrum and its filling by carriers at presence of the CDW with the gap opened in a vicinity of k=±k$_F$ and a pocket of another band protruding the Fermi level near k=0. Right panel: The bottom – the scheme of the mesa-junction with slits cut across the chains' direction x; the black line shows one calculated streamline of the normal current. On top – the calculated 3D plot of the order parameter A (falling from 1 to 0 at points of vorticies' nucleation); the color encodes the electric potential changing the sign from right to left. (Color online)

The GL-like approach assumes that the intrinsic electrons have been integrated out and enter only via the parameters. While such a model works perfectly in static and stationary regimes, the dependence of both the amplitude and the phase of the order parameter on both time and coordinates raises the problem of defining simultaneously the collective charge and the current. Recall that the collective dependencies of the charge $n_c$ and the current $j_c$ in CDW are written commonly as

$$\pi n_c = A^2 \partial_x \varphi \qquad \pi j_c = -A^2 \partial_t \varphi \qquad (1)$$

These definitions are apparent at the temperature $T=0$ when the amplitude saturates to $A=1$. They can be, and have been many times, derived for an elevated *T* when equilibrium constant *A* is getting reduced to *0* above $T_c$. But in a fully spacio-temporal regime we get from (1) the non vanishing rate of the density evolution

$$\pi \frac{dn_c}{dt} = \frac{\partial A^2}{\partial x}\frac{\partial \varphi}{\partial t} - \frac{\partial \varphi}{\partial x}\frac{\partial A^2}{\partial t} \neq 0 \qquad (2)$$

The fictitious particle production appears (2), which is particularly intensive near the moving vortex cores. To resolve the problem, we need to make a step back from integrating out the fermions, considering equations of motions for the order parameter and for all normal carriers separately while in conjugation. The derivation can be done without going to a heavy microscopics, thanks to exploiting [18,19] the chiral invariance and related anomalies. This approach allows taking intrinsic carriers into account explicitly, while at a price of some microscopic constraints: all interactions can depend only on derivatives of the phase $\varphi$ and not on the phase itself. That brings some limitations: the carriers' spectrum is one-dimensional, there are no Umklapp processes due to approaching commensurability, electrons scatter by each other and by CDW phonons [20]. (Some limitations can be released *a posteriori*. Thus the phase dependent scattering by impurities can be included taking into account the counter-current drag: a correction to the normal current proportional to the collective one [21].)

The chiral transformation for wave functions of intrinsic electron $\psi_\pm \rightarrow \psi_\pm \exp(\pm i\varphi/2)$ and $\Psi \rightarrow \Psi \exp(-i\varphi) = \Delta = A\Delta_0$ results in the Hamiltonian where the gap becomes conveniently a real function $\Delta$ but in expense that phase deformation shift the local potential energy and hence effective electric field (the electron charge $e$ is included in $\Phi$ which becomes the energy):

$$\begin{pmatrix} -i\partial_x + \Phi + \partial_x\varphi/2 & \Delta \\ \Delta & i\partial_x + \Phi + \partial_x\varphi/2 \end{pmatrix} \tag{3}$$

$$\Phi \rightarrow \Phi + \frac{\hbar v_F}{2}\frac{\partial \varphi}{\partial x} \quad E \rightarrow E - \frac{\hbar v_F}{2}\frac{\partial^2 \varphi}{\partial x^2} \tag{4}$$

Now we can write the density of the energy functional for all components of the system as

$$C\hbar v_F\left[(\partial_x A)^2 + \alpha(\partial_y A)^2\right] + \hbar v_F/4\pi\left[1 \times (\partial_x \varphi)^2 + \alpha A^2(\partial_y \varphi)^2\right]$$
$$+ 1 \times \Phi \partial_x \varphi/\pi + e\Phi n_{ex} + \left(\Phi + (\hbar v_F/2)\partial_x \varphi\right)n_{in} \tag{5}$$
$$+ F_{CDW}(A, n_{in}, T) + F_{ex}(n_{ex}, T) - (\nabla \Phi)^2 \varepsilon_h s/(8\pi e^2)$$

($C=cnst\sim 1$, $s$ is the area per chain). Here the terms containing derivatives of $A$ are conventional, as well as the terms $\sim \alpha$ - the parameter of the structural anisotropy – which is thought to come from direct interchain interactions, the last term $\sim \varepsilon_h$ - the host dielectric constant - is the electric field energy, $F_{ex}(n_{ex})+e\Phi n_{ex}$ is the total energy of the extrinsic carriers. More particular is the potential energy for intrinsic carriers $(\Phi + \varphi'\hbar v_F/2)n_{in}$ which follows from (3). Also their free energy $F(A, n_{in}, T)$ entangles the dependencies on $n_{in}$ and on $A$, so that its minimum over $A$ depends on $n_{in}$ in such a way that it should disappear when $n_{in}$ exceeds a critical value beyond which the CDW should not be supported. Finally the two terms with $\partial_x\varphi$ - the elastic and the Coulomb energy of condensed electrons of the CDW, look really counterintuitive: their shown factors *1* are expected (in the conventional GL scheme like in [15-17]) to be $A^2$ – at least we expect them vanish together with $A$. The wonder of our approach is that this would happen indeed after elimination of $n_{in}$ from equations of motion generated by this functional. That is proved but cannot be done explicitly rather than with a help of additional approximations.

The system evolution is described by dissipative equations where the LHSs in (6,7,8) are given by the variation of the above functional over $\varphi$, $A$, $\Phi$ correspondingly.

$$\frac{\partial}{\partial x}\left(\frac{\partial \varphi}{\partial x} + \frac{2}{\hbar v_F}\Phi + \pi(n_e - n_h)\right) + \alpha\frac{\partial}{\partial y}\left(A^2\frac{\partial \varphi}{\partial y}\right) = \gamma_\varphi A^2 \frac{\partial \varphi}{\partial t} \tag{6}$$

$$C\nabla^2 A + \alpha A\left(\frac{\partial \varphi}{\partial y}\right)^2 + \frac{\partial F_{CDW}}{\hbar v_F \partial A} = -\gamma_A \frac{\partial A}{\partial t} \tag{7}$$

$$r_0^2 \nabla^2 \Phi + \hbar v_F\left(\partial_x \varphi + \pi(n_{in} + n_{ex})\right) = 0 \tag{8}$$

Here $r_0$ is the Thomas-Fermi screening length for the parent metal. These equations should be completed by the diffusion equations for charge carriers (here we write it for $n_{in}$ only)

$$\nabla(\sigma\nabla\mu) = \frac{e^2}{s}\partial_t n; \quad \mu = \zeta + \Phi + \frac{\hbar v_F}{2}\partial_x\varphi; \quad \zeta = \frac{\partial F_{CDW}}{\partial n_{in}} \tag{9}$$

where $\zeta$ and $\mu$ are the chemical and the electro-chemical potentials, $\sigma=(\sigma_x,\sigma_y)\sim n_{in}$ is the conductivity tensor. In the numerical work we shall assume that extrinsic carriers occupy a 2D small pocket as in Fig.1. The free energy $F(A,n_{in},T)$ and its derivatives over $A$ in (8) and over $n_{in}$ in (9) are calculated numerically from thermodynamics of the CDW state and then interpolated by plausible analytic expressions.

Equations (6)-(9) must be supplemented by boundary conditions. The CDW stress (derivatives of the energy (5) over $\partial_i\varphi$) normal to the boundary should vanish. The normal electric field will be chosen also as zero at all boundaries which ensures the total electro-neutrality and confines the electric potential within the sample which greatly simplifies calculations. There should be no normal current flow at the boundaries except for the two source/drain boundaries left for the applied voltage.

## 3. Modeling results.

A numerical solution of eqs. (6-9) was performed by the finite element method using the program environment of COMSOL. We tried the full set of eqs. as well as an approximation of very high normal conductivity when the eq. (9) is reduced to the condition $\mu=cnst$. Another possible approximation is the local electro-neutrality condition when $r_0=0$ ($r_0 \sim$nm is always small indeed) in the Poisson eq. (8). Unlike the former work on much more easy GL eqs., here we studied mostly the simplified rectangular geometry of the junction with the voltage applied at opposite boundaries in the inter-chain direction y (see Fig.2). This rectangle is thought to be isolated as the bridge area among overlapping slits in Fig.1 – right bottom drawing; here the current is forced in the transverse direction of the high resistance and the major voltage drop takes place.
A priori we faced a strong shape effect: the vortices nucleate, and finally one of them fully develops, in all four sample's corners as in Fig.2-left. While that may have significance and actually take place, we are interested in more universal features; recall that the rectangle is a bit artificial cut from the true experimental geometry of the mesa junction made by the slits. Since we could not go to very long samples for computational reasons, we have just applied the conditions $A = 1$ at the short-side boundaries, thus preserving there the nominal CDW amplitude. The two-vortex configurations in Fig.2 has been computed for different inter-chain coupling and on-chain conductivity. In case of a weak inter-chain coupling and higher on-chain conductivity
The middle panel gives the moment when one vortex has reached the full nucleation while another one is still in development. The effect of vortices is profound suppressing the order parameter across the whole width. That was calculated from the full system of eqs. for a high ~1000 anisotropy of conductivities. The right panel gives the simultaneous nucleation of a pair of vortices at opposite boundaries. That was calculated in the approximation $\mu=cnst$ which follows from eq. (9) when both components of the conductivity are equally infinity so the relaxation of electrons is instantaneous.

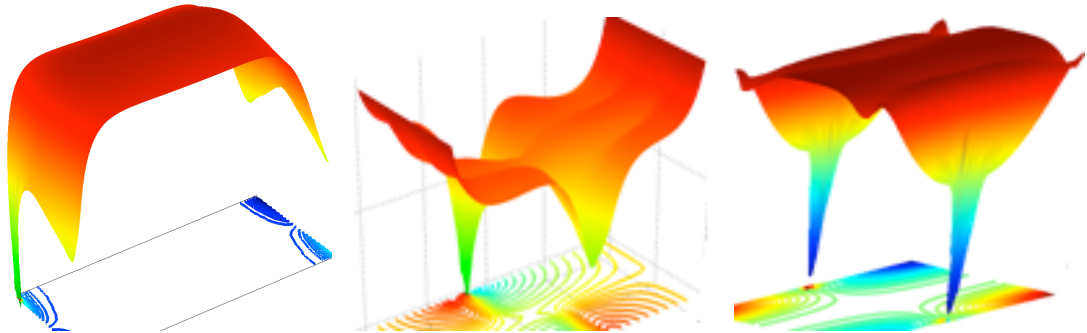

Figure 2. 3D surfaces of the order parameter amplitude and corresponding contour plot for the phase. Left: unrestricted solution of the full eqs. system; Middle: restrictions preventing corner vortices. Right: also restricted, approximate eqs.

Unlike the earlier simpler work [15-17], now we could not follow the evolution of vortices after their nucleation at boundaries, when they are expected to proliferate inwards the bulk [11,12]. Now that is prevented by the program instability which is explained by the nonanalytic dependences of new eqs. on the order parameter. There is no explicit compensation of diverging $\partial_x \varphi$ by vanishing $A^2$. This difficulty can be overcome and the detaching of vortices was seen indeed by means of the more problem-specific programming. But the calculations become annoyingly slow, taking days for each run, so the results will be reported elsewhere when the full set of data is ready for analysis [22].

## 4. Conclusion

We have derived the extended GL theory for the CDW which allows taking intrinsic carriers explicitly into account. The derivation is based on the property of the chiral invariance and takes into account the chiral anomaly. The resulting equations are qualitatively more complicated than what is naturally assumed and was exploited in earlier studies. Equations rely on hidden compensations of divergences which should rise at the vortex cores. That makes them not analytic in the order parameter which brings computational problems in the numerical modeling. Even if we were not able to reach the regime of the vortex multiplication, propagation and annihilation as it was seen spectacularly in [15-17], still we can observe their nucleation at boundaries or at obstacles. More sophisticated deducted programming is in progress. Further extension of theory is necessary to take into account the current conversion effect: exchange or electrons among the reservoirs of condensed and excited particles. That requires for more carful distinction of the glide and the climb (motions along and transversely to chains direction) [23].

The studied reconstruction in junctions of the CDW can be relevant to modern efforts of the field-effect transformations in strongly correlated material which also show a spontaneous symmetry breaking [24]. The most recent progress was a switching under an applied voltage among different states of an electronic crystal in 1T-TaS2 [25].